\documentclass[twocolumn,trackchanges]{aastex701}
\usepackage{amsmath}
\usepackage{graphicx}
\usepackage{dcolumn}
\usepackage{bm}
\usepackage{xcolor}
\usepackage{comment}
\usepackage{hyperref}

\begin{document}

\title{Distributions of particles accelerated by strong Alfv\'enic turbulence}

\correspondingauthor{Stanislav Boldyrev}
\email{boldyrev@wisc.edu}

\author[orcid=0000-0001-6252-5169]{Stanislav Boldyrev}
\affiliation{Department of Physics, University of Wisconsin at Madison, Madison, Wisconsin 53706, USA}
\affiliation{Center for Space Plasma Physics, Space Science Institute, Boulder, Colorado 80301, USA}
\email{}

\author{Daniel Humphrey}
\affiliation{Department of Physics, University of Wisconsin at Madison, Madison, Wisconsin 53706, USA}
\email{}

\author[orcid=0000-0003-1745-7587]{Vadim Roytershteyn}
\affiliation{Center for Space Plasma Physics, Space Science Institute, Boulder, Colorado 80301, USA}
\affiliation{Los Alamos National Laboratory, Los Alamos, New Mexico 87545, USA}
\email{}

\author{Cristian Vega}
\affiliation{Department of Physics, University of Wisconsin at Madison, Madison, Wisconsin 53706, USA}
\email{}

\begin{abstract}
This work presents a model for generating nonthermal power-law tails of particles' energy probability density functions in turbulent collisionless plasmas, applicable to both non-relativistic and relativistic scenarios. We propose that strong Alfvénic turbulence energizes plasma particles through curvature acceleration, particularly for particles with Larmor radii comparable to the scales of turbulence. When the energy density of the energized particles increases, the efficiency of the energy exchange process
diminishes. As a result, the acceleration process saturates, leading to power-law distributions of particle momentum and energy. In the non-relativistic case, the momentum probability density function scales as  $f(p) dp \propto p^{-3} dp $, while in the ultrarelativistic case, the energy probability density function scales as $ f(\gamma) d\gamma \propto \gamma^{-3} d\gamma $, where $\gamma$ is the Lorentz factor. This model provides a unified framework for understanding particle acceleration in both energy regimes, complementing existing analytical approaches.  The predicted scalings are consistent with available observations of energetic ion distributions in the heliosphere and with the findings from numerical simulations of ultrarelativistic particle acceleration in magnetically dominated plasma turbulence.
\end{abstract}



\section{Introduction}
Power-law distributions of energetic suprathermal particles are observed in various space and astrophysical environments, ranging from non-relativistic solar wind plasmas to high-energy astrophysical systems where plasmas reach ultrarelativistic temperatures. Some of the most familiar examples include the so-called ``halo" ions in the inner and outer heliosphere \cite[e.g.,][]{gloeckler1992,gloeckler2008,mason_g2012,fisk2014,kolmann2019}, and the power-law spectra of ultrarelativistic electrons, which are inferred from observations of synchrotron radiation emitted by pulsar wind nebulae \cite[e.g.,][]{atoyan1996,meyer2010,abdo2011a,abdo2011b}. Both in situ observations of the solar wind and numerical simulations of relativistic and non-relativistic systems suggest that turbulence may play a crucial role in accelerating particles to non-thermal energies \cite[e.g.,][]{sironi2014,marcowith2016,servidio2016,trotta2020,ergun2020a,lemoine2021,lemoine2024,pezzi2022,bresci2022,sironi2022,dong2022,french2023,vega2020,vega2022a,xu2023,das2025,wong2025,grosselj2026}. A variety of analytical approaches have been developed to address power-law energy distribution functions in non-equilibrium plasmas \cite[e.g.,][]{bykov2001,drake2013,fisk2014,vega2022a,zhdankin2022,banik2024,ewart2025,meringolo2026}. 

In a collisionless plasma, a particle's magnetic moment is conserved unless strong resonance interactions occur with turbulent fluctuations. This conservation of magnetic moment enables a particle to interact effectively with structures, which may contribute to nonlinear interactions that play a role in the acceleration of suprathermal particles. Recent analytical and numerical studies of relativistic magnetically dominated turbulence indicate that curvature acceleration can significantly influence particle energization \cite[e.g.,][]{lemoine2021,vega2024b,sebastian2025}. Indeed, because of the critical balance and local anisotropy of Alfvénic turbulent fluctuations, particles that interact most effectively with the turbulent eddies tend to have small pitch angles. In this case, the primary drift leading to energy exchange between particles and turbulent fields is the curvature drift.

In this work, we propose a model for generating nonthermal power-law tails in a turbulent collisionless plasma, applicable to both non-relativistic and relativistic scenarios. We assume that strong Alfvénic turbulence can effectively energize plasma particles with Larmor radii comparable to the scales of turbulent fluctuations through the process of curvature acceleration. However, as the energy density of energized particles increases, the efficiency of the energy exchange process diminishes. Consequently, the particle probability density functions for energy and momentum can saturate, resulting in power-law shapes.

Our  model predicts that in the non-relativistic case, the momentum probability density function scales as $ f(p) dp \propto p^{-3} dp $. In the ultrarelativistic case, the energy probability density function, expressed in terms of the particle's Lorentz factor, scales as $ f(\gamma) d\gamma \propto \gamma^{-3} d\gamma $. These phenomenological predictions are consistent with the nonthermal distributions of energetic particles found in the heliosphere and those obtained through numerical simulations. Our consideration offers a unified treatment of both non-relativistic and relativistic cases of high and low magnetization, and is complementary to previous analytical approaches.

In the following sections, we will first formulate our phenomenological model for the non-relativistic case and detail the key assumptions underlying the derivation. We will then extend this model to other cases, including turbulence in a relativistic plasma, addressing both magnetically dominated and non-magnetically dominated regimes.

\section{Non-relativistic case}
\label{nonrel_model}
Let us represent the spectrum of magnetic fluctuations in the inertial interval of Alfv\'enic turbulence as
\begin{eqnarray}
\label{turb_energy}
{\cal E}_B(k_\|)\,dk_\|={A_B}{k_\|^{-2}}\,dk_\|, 
\end{eqnarray}
where $A_B= (\delta B_0)^2 /(8\pi L_\|)$ is the normalization coefficient, $L_\|$~ is the outer scale of turbulent fluctuation along the uniform guide field, $B_0$ is the strength of the guide field, $\delta B_0$ is the typical strength of magnetic fluctuations at the outer scale, and $k_\|$ is the wave number of the turbulent fluctuations in the {\em local} direction of the magnetic field line.\footnote{{To derive this formula, we note that the intensity of magnetic fluctuations at the field-parallel scale $l$ can be defined as 
\begin{align*}
\delta B^2(l)&=\frac{1}{2}\langle\left[\delta {\bf B}(z+l)-\delta{\bf B}(z)\right]^2 \rangle \\
&=\int_0^\infty I(k_\|)(1-e^{ik_\|l})dk_\|\approx \int_{1/l}^\infty I(k_\|)dk_\|,
\end{align*}
while the energy density of magnetic fluctuations is
\begin{align*}
\frac{1}{8\pi}\langle[\delta {\bf B}(z)]^2\rangle=\frac{1}{8\pi}\int_0^\infty I(k_\|)dk_\|=\int_0^\infty {\cal E}_B(k_\|)dk_\|.
\end{align*}
The energy of magnetic fluctuations at the outer scale $L_\|$ is then:
\begin{align*}
\frac{1}{8\pi}(\delta B_0)^2\equiv \frac{1}{8\pi}\delta B^2(L_\|)=\int_{1/L_\|}^\infty {\cal E}_B(k_\|)dk_\|.
\end{align*}
}} This spectrum is consistent with phenomenological considerations of strong Alfv\'enic turbulence~\cite[e.g.,][]{goldreich_toward_1995,boldyrev2006,chandran_15}.

We assume that the plasma particles probability density function in momentum space, $f(p)$, is normalized such that $\int f(p) \, dp = 1$. In the ultra-relativistic case, it is more convenient to consider the distribution of the particle's Lorentz factors, $ f(\gamma) $, normalized as $\int f(\gamma) d\gamma = 1 $.  

For our discussion, it is essential to notice that energetic particles can interact with turbulent fluctuations strongly, in the sense that such interaction may significantly alter the energy of the turbulent fluctuations in a single interaction. When an energetic particle with momentum $p$ propagates along a curved magnetic field line generated by Alfv\'enic turbulence, it experiences the curvature of the eddies whose field-perpendicular size is at least as large as particle's field-perpendicular gyroradius. As discussed in \cite[][]{vega2024b,vega2025,humphrey2026}, this means that the smallest eddy that can guide such a particle, has the field parallel size of~$l\sim p/(\Omega m)$, where $\Omega=q B_0/(m c)$, and $m$ is the particle's mass. This condition means that the smallest eddy that can nonlinearly exchange energy with the particle due to curvature acceleration, should formally be in cyclotron resonance with it, 
\begin{eqnarray}
\label{resonance}
k_\|p =\pm \Omega m, 
\end{eqnarray}
where $k_\|=2\pi/l$. The results discussed in this paragraph hold for both non-relativistic and relativistic cases; in the latter case, one needs to use the relativistic expression for the particle's momentum,~$p =\gamma m v $.

In the non-relativistic case, a particle interacts with the Alfv\'enic eddy if the eddy does not decay due to nonlinear interactions while the particle propagates along it. For that, we need to additionally require that the velocity of the energetic particle exceeds the Alfv\'en velocity, $p/m \gtrsim v_A$. In a relativistic magnetically dominated plasma, both velocities are close to the speed of light, so this condition is automatically satisfied.

The condition given by Eq.~(\ref{resonance}) implies, quite importantly, that the pitch angle of the particle interacting with an eddy of parallel size $l$ coincides with the anisotropy angle of the eddy itself, $\theta\sim k_\|/k_\perp \ll 1$. 
In Alfv\'enic turbulence, one evaluates this angle as\footnote{This formula follows from the critical balance condition $k_\|/k_\perp \approx \delta B(l)/B_0$, and from the field-parallel scaling of magnetic fluctuations in Alfv\'enic turbulence, $\delta B(l)\approx (l/L_\|)^{1/2}\delta B_0$.}:
\begin{eqnarray}
\label{angle}
\theta^2\approx \frac{(\delta B_0)^2}{B_0^2}\frac{l}{L_\|}=\frac{(\delta B_0)^2}{B_0^2}\frac{2\pi p}{m\Omega L_\|}.
\end{eqnarray}
The fraction of the solid angle occupied by particles interacting with such an eddy is $\Delta \Omega/4\pi=1-\cos(\theta)\approx \theta^2/2$, where we assume that the angle is small and particles can propagate in both directions. This formula is applicable in both relativistic and non-relativistic cases. 

Recent studies have identified curvature drift as a significant factor in the acceleration of energetic particles \cite[e.g.,][]{lemoine2023b, vega2024b,sebastian2025}. Indeed, according to Eqs.~(\ref{resonance}) and~(\ref{angle}), the field-perpendicular gyroradii of particles that interact with anisotropic turbulent eddies must be relatively small. Since curvature drift is the only type of drift that 
survives in the limit of small field-perpendicular gyroradius, it dominates over other drifts during interactions at small pitch angles. In this discussion, we will explore the energization caused by curvature drift in greater detail. 

Consider a turbulent eddy in Alfvénic turbulence that has a field-parallel size of~$l$. The average magnetic field line curvature in such an eddy is estimated as $1/R_c=(4/l) \delta B(l)/B_0$, while the velocity fluctuation associated with the eddy is $v(l)\approx v_A\delta B(l)/B_0$. After the particle has propagated through such an eddy, its energy change due to the curvature drift is then estimated as
\begin{eqnarray}
\label{delta_E}
\Delta {\cal E}\approx 4 p v_A\frac{(\delta B_0)^2}{B_0^2}\frac{l}{L_\|},    
\end{eqnarray}
see, e.g., \cite[][Eq.~(5)]{northrop1963,vega2024b}. This formula applies to both relativistic and non-relativistic scenarios. In the ultrarelativistic case, the particle's momentum should be expressed as $p \approx \gamma mc $. In the magnetically dominated case, the Alfvén speed~$v_A$ should additionally be replaced with the speed of light~$c$.

During a single crossing time, the energy density of turbulent fluctuations at scale~$l$ interacting with energetic particles, can be altered due to this interaction by an amount:
\begin{eqnarray}
\label{E_change}
n\Delta {\cal E}\frac{\theta^2}{2}f(p)\,dp,
\end{eqnarray}
where $n$ is the average density of the particles. Since the time it takes for a particle to cross a turbulent eddy is shorter than or similar to the eddy turnover time (the time it takes for energy to be replenished in the eddy through the energy cascade), a strong interaction can significantly impact the eddy's energy. This effect occurs when the energy being transferred is comparable to the energy present in the turbulent fluctuations at the corresponding scales, as defined by Eq.~(\ref{turb_energy}).

Comparing the two energy densities, we obtain {the condition for such a ``dynamic balance"}:
\begin{eqnarray}
\label{E_balance}
 {A_B}{k_\|^{-2}}\,dk_\| \approx  n\Delta {\cal E}\frac{\theta^2}{2}f(p)\,dp.  
\end{eqnarray}
Noting that $l=2\pi/k_\|$,  $k_\|=\Omega m/p$, and $k_\|^{-2}dk_\|=-dp/(\Omega m)$, we can derive from this equation the asymptotic form of the distribution function at large momenta. Restricting ourselves to the nonrelativistic case and the ion-electron plasma, we can derive the asymptotic boundary for the ion distribution function: 
\begin{eqnarray}
\label{f_asympt}
f_i(p)dp\approx \frac{B_0^2}{(\delta B_0)^2}\frac{L_\|}{128\pi^3d_i}\frac{m_i^2v^2_A}{p^5}\,4\pi p^2 dp.\,
\end{eqnarray}
Here, we use the nonrelativistic Alfv\'en velocity, $v_A=B_0/\sqrt{4\pi n m_i}$, and the ion inertial scale $d_i=v_A/\Omega_i$. 

In this model, the energetic particles get accelerated by turbulent fluctuations until they reach asymptotic distribution~(\ref{f_asympt}). In order to be normalizable, such a distribution function can form only at 
\begin{eqnarray}
v\gg v_A\sqrt{\frac{B_0^2}{(\delta B_0)^2}\frac{L_\|}{64\pi^2 d_i}}
\end{eqnarray}
This power-law tail will, therefore, be especially pronounced in the regions where the turbulence is stronger, the outer scale of turbulence, $L_\|$, is smaller, and the resulting curvature of the field lines is larger.

\section{Non-relativistic case: Equipartition hypothesis}
\label{section3}
 Here, we discuss a modification of the previous model, where we do not assume a specific form for the nonlinear interaction mechanism. This approach may be reasonable when no single interaction leads to significant energy exchange. Instead, we assume that decaying turbulence tends to achieve energy equipartition with the accelerated particles at each scale. This assumption does, however, require further justification, as interactions that do not significantly alter eddy energy within one crossing time can allow eddy energy to be quickly replenished by the energy cascade. Therefore, one may not, in general, expect energy equipartition to occur. However, we present this discussion here because it formally leads to the same scaling of the particle momentum distribution function as in Section~\ref{nonrel_model}, although with a different dependence on the plasma and turbulence parameters. Consequently, it is worth comparing this equipartition model with the more physically motivated model of Section~\ref{nonrel_model}.
 
As in the previous section, a particle most efficiently interacts with turbulence when it propagates within a small angle $\theta\approx k_\|/k_\perp$ with respect to the local magnetic field line. We assume that the velocities of accelerated particles exceed the Alfv\'en velocity, and, therefore,  $k_\|v\gg \omega$. The cyclotron wave-particle resonance condition is then given by a simple relation, $k_\|v= \pm\Omega_i$, and we can evaluate this angle as:
\begin{eqnarray}
\label{angleB}
\theta^2\approx \frac{(\delta B_0)^2}{B_0^2}\frac{l}{L_\|}= \frac{(\delta B_0)^2}{B_0^2}\frac{2\pi v}{\Omega_i L_\|}.
\end{eqnarray}
The fraction of the solid angle occupied by such particles, propagating in both directions, is given by the same expression as in Section~\ref{nonrel_model}, $\Delta \Omega/4\pi=1-\cos(\theta)\approx \theta^2/2$, where the angle is given by Eq.~(\ref{angle}). 

The energy density of such particles is given by
\begin{eqnarray}
\frac{n p^2}{2m_i}\frac{\theta^2}{2}f_i(p)\,dp.
\end{eqnarray}
We can equate this energy to the energy contained in the turbulent fluctuations at the corresponding scales, given by Eq.~(\ref{turb_energy}), to obtain:
\begin{eqnarray}
 {A_B}{k_\|^{-2}}\,dk_\| \approx  \frac{n p^2}{2m_i}\frac{(\delta B_0)^2}{B_0^2}\frac{2\pi v}{\Omega_i L_\|}f_i(p)\,dp.  
\end{eqnarray}
By using $k_\|^{-2}dk_\|=-dp/(\Omega_im_i)$, we derive from this equation the asymptotic boundary for the probability density function at large momenta, 
\begin{eqnarray}
\label{f_asymptB}
f_i(p)dp\approx \frac{1}{4\pi^2}\frac{m_i^2v^2_A}{p^5}\,4\pi p^2 dp.
\end{eqnarray}
Remarkably, this distribution is quite universal. Although it depends on the presence of Alfv\'enic turbulence and the strength of the guiding field, it remains independent of the turbulence intensity, $(\delta B_0/B_0)^2$.  In order to be normalizable, this function can form only at 
\begin{eqnarray}
v\gg \frac{v_A}{\sqrt{{8\pi^2}}}.    
\end{eqnarray}
Comparing these analytical predictions with observational data may help verify the applicability of the equipartition model to describe nonthermal particles in specific cases.

\section{Relativistic magnetically dominated case}
\label{section4}
A typical setup for studying such a case is to consider relaxation of magnetic fluctuations in an initially magnetically dominated plasma, where the energy contained in the initial magnetic fluctuations significantly exceeds the rest mass energy of the plasma particles. Since in such a set up, plasma is rapidly heated to ultrarelativistic temperatures, we consider for simplicity the case of electron-positron pair plasma.  We denote the magnetization parameter based on magnetic fluctuations as:
\begin{eqnarray}
{\tilde \sigma}_0=\frac{(\delta B_0)^2}{4\pi n m_e c^2},    
\end{eqnarray}
while the magnetization parameter based on the uniform part of the field (the guide field) is
\begin{eqnarray}
\sigma_0=\frac{B_0^2}{4\pi n m_e c^2}. 
\end{eqnarray}

In the magnetically dominated case, both magnetization parameters are large, ${\tilde \sigma}_0\gg 1$, and $\sigma_0\gg 1$. 

Similarly to what is done in some numerical setups \cite[e.g.,][]{comisso2018,comisso2019,vega2022a,vega2023,vega2024b,vega2025,sebastian2025,humphrey2026}, we assume that the plasma initially is mildly relativistic, $k_B T_0\sim mc^2$.  However, when the magnetic energy is transferred to plasma particles, the resulting particle energies will inevitably become ultra-relativistic. The particle heating continues until the particle thermal energy becomes comparable to the (declining) energy of electromagnetic fluctuations.   The resulting particle thermal energy will therefore be approximately equal to a half of the initial magnetic energy, corresponding to the approximate ``thermal" gamma factor of $\gamma_{th}\sim {\tilde \sigma}_0/4$. By the time the magnetic fluctuations have significantly decreased and thermalized with the particles, the acceleration processes will mostly be completed, leading to the formation of power-law distributions of the energetic particles.

Strong Alfv\'enic turbulence in an ultrarelativistic plasma shares similarities with its non-relativistic counterpart. The Alfvénic velocity fluctuations are limited by the sound speed, which is $c/\sqrt{3}$, making them at most mildly relativistic \cite[e.g.,][]{zhdankin2017a,vega2022a}. The spectrum of these fluctuations and their anisotropy display similar characteristics between the relativistic and nonrelativistic cases \cite[e.g.,][]{zhdankin2017a,chernoglazov2021,vega2022b,nattila2020,nattila2022}.  The equations for the curvature drift acceleration (\ref{delta_E}) and the anisotropic angle (\ref{angle}) can also be easily adapted for the relativistic case. Here we present them for convenience, where we have also expressed the energy change in Eq.~(\ref{delta_E}) through the gamma factor of an ultrarelativistic particle:
\begin{eqnarray}
\label{angle2}
\theta^2\approx \frac{(\delta B_0)^2}{B_0^2}\frac{l}{L_\|}= \frac{(\delta B_0)^2}{B_0^2}\frac{2\pi\gamma c }{L_\|},
\end{eqnarray}
\begin{eqnarray}
\label{delta_E2}
\Delta {\cal E}\approx 4 \gamma m_ec^2\frac{(\delta B_0)^2}{B_0^2}\frac{l}{L_\|}.   
\end{eqnarray}

Substituting these formulae into the energy balance Eq.~(\ref{E_balance}), where we need to add the equal contributions of the electrons and positrons, we obtain the limiting asymptotic form of the particle distribution function at large energies. Here, we express this distribution through the distribution of the $\gamma$-factors:
\begin{eqnarray}
\label{f_rel}
f(\gamma)d\gamma\approx \frac{\sigma_0}{32\pi^2}\frac{B_0^2}{(\delta B_0)^2}\frac{\gamma_{max}}{\gamma^3}\,d\gamma.    
\end{eqnarray}
In this formula, we have denoted by $\gamma_{max}=L_\| \Omega_e/c$ the Lorentz gamma factor at which the electron gyroradius becomes comparable to the outer scale of turbulence. In order to be normalizable, this asymptotic tail can form only at 
\begin{eqnarray}
\gamma \gg \sqrt{\frac{B_0^2}{(\delta B_0)^2}\frac{\sigma_0 \gamma_{max}}{64 \pi^2}}.     
\end{eqnarray}
{~}\\

\section{Relativistic case with low magnetization}
\label{section5}
We now consider the case when the plasma has ultrarelativistic temperature and low magnetization, $\sigma=B_0^2/(4\pi w nmc^2)\ll 1$, where $w$~is the plasma enthalpy per particle. Such cases were considered, e.g., in numerical simulations of driven turbulence in a pair plasma in~\cite[][]{zhdankin2020,wong2025}, when the large-scale driving force maintained the $\delta B_0/B_0\sim 1$ level of turbulence over many turnover times. As a consequence, the force continually supplied energy to the plasma, heating it to ultrarelativistic temperatures and causing a gradual decrease in magnetization~$\sigma$.

Formulae (\ref{delta_E}), (\ref{E_change}), and (\ref{E_balance}) are valid in this case, where we have $p=\gamma mc$, and $v_A/c\approx\sqrt{\sigma}\ll 1 $. Instead of Eq.~(\ref{f_rel}), we now have:
\begin{eqnarray}
\label{f_rel2}
f(\gamma)d\gamma\approx \frac{c}{v_A}\frac{\sigma_0}{32\pi^2}\frac{B_0^2}{(\delta B_0)^2}\frac{\gamma_{max}}{\gamma^3}\,d\gamma,
\end{eqnarray}    
which is applicable if
 \begin{eqnarray}
\gamma \gg \sqrt{\frac{c}{v_A}\frac{B_0^2}{(\delta B_0)^2}\frac{\sigma_0 \gamma_{max}}{64 \pi^2}}.     
\end{eqnarray}   
{~}\\

\section{Relativistic equipartition model}
Finally, we consider the formal equipartition model for the relativistic case, which can be derived analogously to the model of Section~\ref{section3}.  In this case, the anisotropy of turbulent fluctuations is given by Eq.~(\ref{angle2}), so the energy equipartition condition reads:
\begin{eqnarray}
2n\gamma m_ec^2\frac{(\delta B_0)^2}{2 B_0^2}\frac{l}{L_\|}f(\gamma)d\gamma \approx \frac{(\delta B_0)^2}{8\pi L_\|}\frac{dk_\|}{k_\|^2}. \,\,  
\end{eqnarray}
The resulting limiting distribution is then
\begin{eqnarray}
\label{equip}
f(\gamma)\approx \frac{\sigma_0}{4\pi\gamma^2},   
\end{eqnarray}
which is applicable at $\gamma\gg\sigma_0/4\pi$. 

Similarly to the nonrelativistic case discussed in Section~\ref{section3}, this shape remains unaffected by the intensity of turbulent fluctuations and the level of magnetization,~${\tilde \sigma}_0$. Although the predicted energy scaling of $\gamma^{-2}$ does not agree with numerical simulations in regimes where $\delta B_0/B_0 \sim 1$, this prediction may hold true for two-dimensional cases of decaying turbulence where the guide field is weak,~$\delta B_0/B_0 \gg 1$. Indeed, in such cases, in the limit of strong magnetization ${\tilde \sigma}_0\gg 1$, the numerically obtained energy spectra approach the scalings close to $\gamma^{-2}$ \cite[e.g.,][]{comisso2018,vega2022a}. 

{The $-2$ spectral exponent for particle energy distribution functions has been predicted in previous models that examined the equilibrium between particles and turbulence \cite[][]{lemoine2024, ewart2025, mbarek2026}. A study closely related to our discussion is the recent work by \citet{mbarek2026}, which derived a formula similar to Eq.~(\ref{equip}) and also investigated how intermittency effects may lead the equilibrium energy spectrum to steepen to values below~$-2$.}
{~}\\

\section{Discussion and conclusion}
{We have proposed a phenomenological model that examines the distribution of energetic particles resulting from a scale-by-scale dynamic balance. This balance occurs between the energy transferred from an Alfv\'enic eddy to energetic particles and the energy supplied to the eddy by a turbulent cascade.} We assumed that the primary process of energy exchange between particles and turbulence is driven by the curvature drifts of particles and the associated curvature acceleration in the magnetic fields produced by anisotropic Alfvénic eddies. In the non-relativistic case, the model predicts the limiting particle momentum distribution function's scaling $f(p)dp\propto p^{-3}dp$, while in the ultrarelativistic case, the same scaling holds for the energy distribution function $f(\gamma)d\gamma\propto \gamma^{-3}d\gamma$. These scaling laws are consistent with the nonthermal distributions of energetic particles found in the heliosphere and those obtained through numerical simulations \cite[e.g.,][]{gloeckler2008,mason_g2012,kolmann2019,fisk2014,zhdankin2017a,comisso2018,vega2022a,vega2024b,vega2025,wong2025,humphrey2026}.

\begin{figure}[h!]
\includegraphics[width=1.\columnwidth]{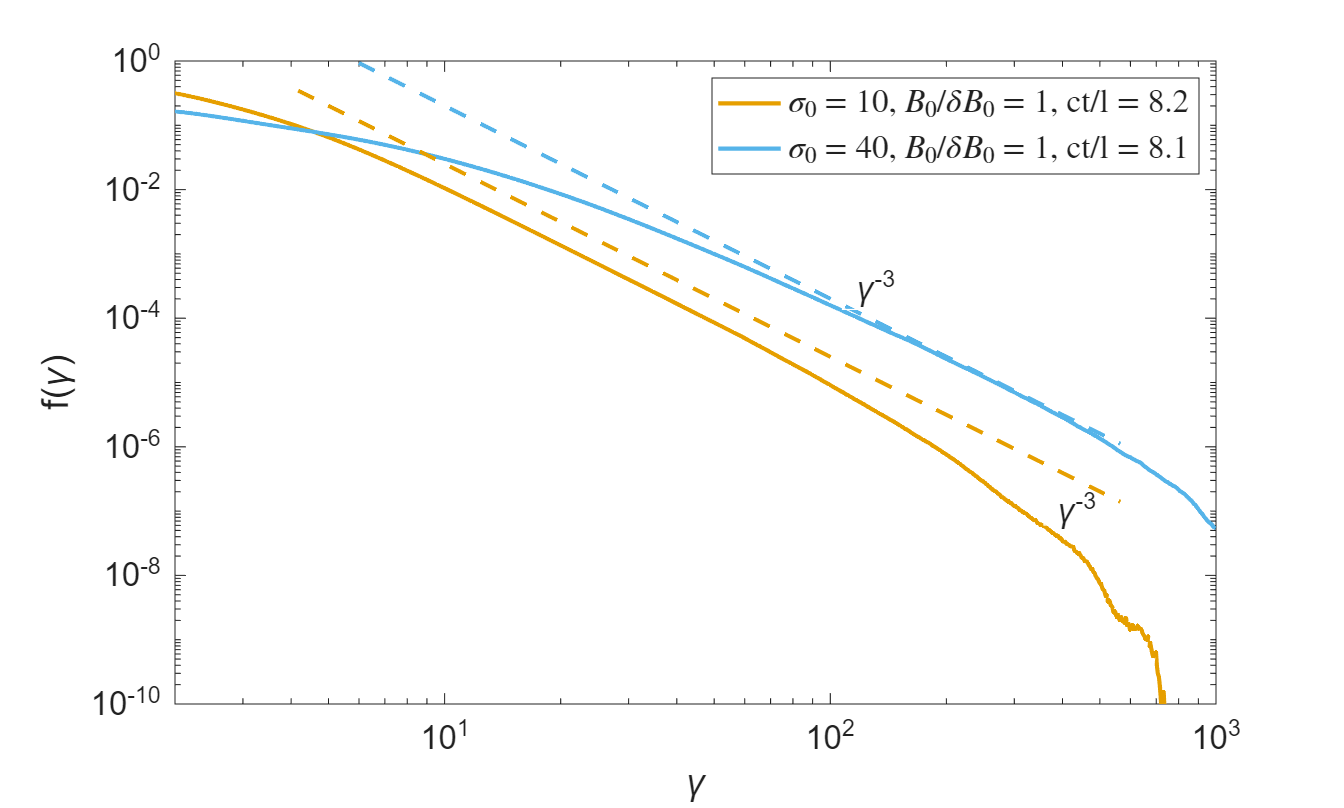}
\hskip-1cm \includegraphics[width=1.\columnwidth]{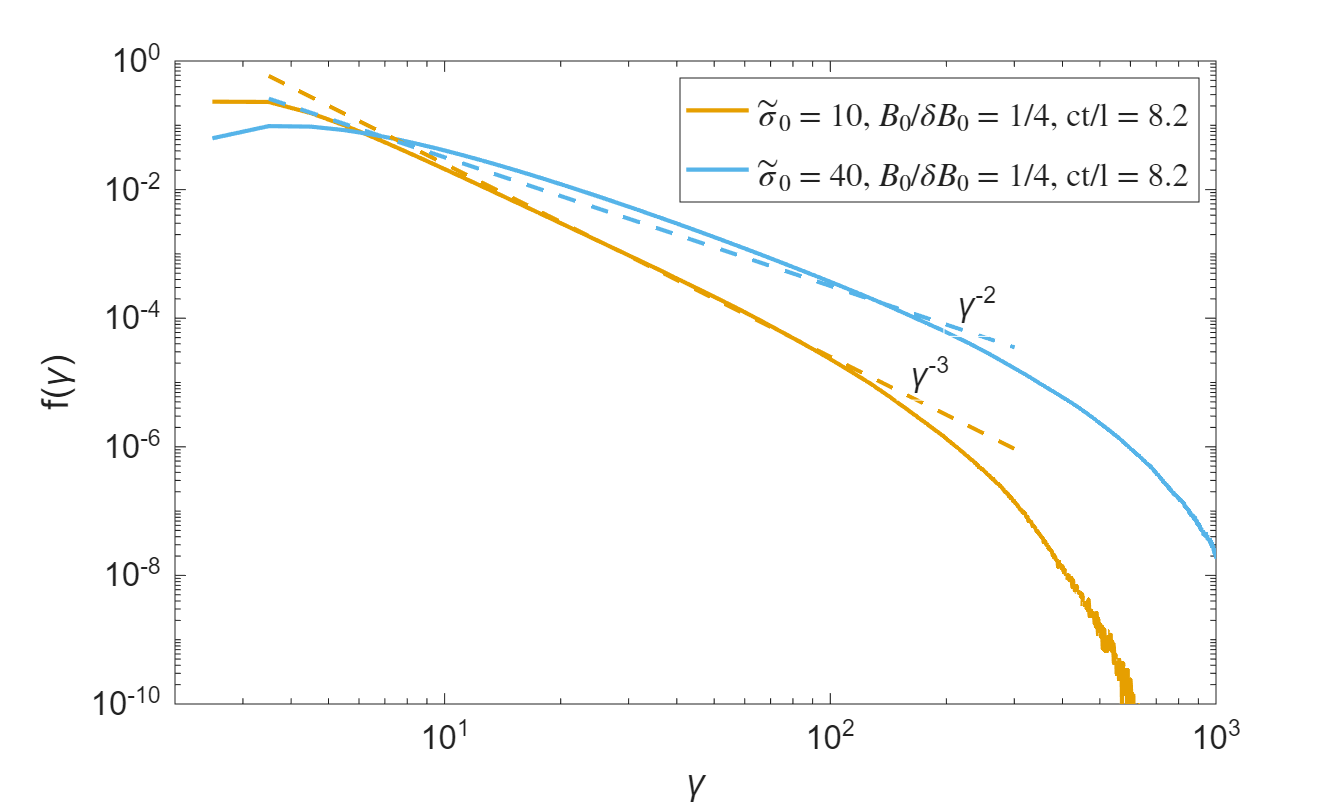}
\caption{{Particle energy probability density functions obtained from ``2.5D" particle-in-cell (PIC) simulations of decaying magnetically dominated turbulence for varying magnetization parameters and strengths of the guide field. Details of the simulations can be found in \citet{vega2022a}. The turbulence was initialized in a square box $ L \times L = 2010 d_e \times 2010 d_e $, using $ N = 8 $ first Fourier modes. The outer scale of the turbulence is then defined as $L_{out} = L/8 $.
In the top panel, we calculate the maximum gamma factor as $ \gamma_{max} = L_{out} \sqrt{\sigma_0}/d_e $, with $ B_0/{\delta B}_0 = 1 $. In the bottom panel, the role of the guide field is played by large-scale turbulent fluctuations, allowing us to effectively assume $ B_0/{\delta B}_0 \approx 1 $ and use the definition \( \gamma_{max} = L_{out} \sqrt{{\tilde \sigma}_0}/d_e \).}} 
\label{vega_example}
\end{figure}

{We have also presented models that assume scale-by-scale energy equipartition of particles with turbulence fluctuations, and we compared and contrasted the predictions of these models with those based on dynamic balance.}

It is interesting to compare our predictions in more detail with numerical simulations of magnetically dominated turbulence. In Figure~\ref{vega_example}, we include plots from the paper by \cite[][]{vega2022a}. The dashed lines represent the asymptotic power-law distributions calculated based on our formulae (\ref{f_rel}, \ref{equip}), and the parameters of the corresponding runs. The agreement between our predictions and the simulations is reasonably good. The discrepancies  between the measured distribution functions and the curves predicted by our model may stem from the phenomenological model's limitations in accurately predicting numerical coefficients.

We can also discuss possible modifications to the model. Our model equations $(\ref{f_asympt})$, $(\ref{f_rel})$, and (\ref{f_rel2}) predicts a linear dependence of the saturated function on the outer scale of turbulent fluctuations. However, this may not align with the results of available simulations, prompting further refinements. One potential modification is to take into account that numerical observations—such as those described in e.g.,~\cite[][]{vega2023}—indicate that the fraction of the volume occupied by accelerated particles is typically smaller than the simulation volume. This suggests that regions of strong turbulence are not space-filling. Particles are predominantly accelerated in regions of enhanced turbulence that are associated with strong current sheets and reconnection events. 

It is reasonable, therefore, to assume that the thickness of such regions scales with the reconnection scale, which in a relativistic pair plasma would be the relativistic electron-inertial scale, $\sim d_{rel}$. The other two linear dimensions are, however, associated with the large scales of the turbulence. Consequently, the volume fraction of regions with enhanced turbulence, where particles are primarily accelerated, scales inversely with the outer scale of turbulence. Our formulas $(\ref{f_asympt})$, $(\ref{f_rel})$, and (\ref{f_rel2}) are applicable to the regions where particles are accelerated. However, in order to apply them to the entire turbulent domain, it is necessary to include the volume filling factor of such regions, which scales proportionally to ${d_{rel}}/{L_\|}$. This would naturally reduce the dependence of the fraction of the accelerated particles on the outer scale of turbulence, and would be in agreement with numerical simulations. The actual values of the volume filling factor may, however, differ between the relativistic and non-relativistic cases, as well as between magnetically dominated and non-magnetically dominated regimes. Future numerical studies of three-dimensional cases or, where possible, observational data with varying parameters may help to refine this phenomenological model.
{~}\\

\begin{acknowledgments}
 This work was supported by the U.S. Department of Energy, Office of Science, Office of Fusion Energy Sciences under award number DE-SC0024362. The work of D.H. and S.B. was also supported by the University of Wisconsin-Madison, Office of the Vice Chancellor for Research, with funding from the Wisconsin Alumni Research Foundation. S.B. acknowledges the hospitality of the KITP program ``Relativistic Plasma Physics: From the Lab to the Cosmos," where this work was initiated. It was supported in part by grant NSF PHY-2309135 to the Kavli Institute for Theoretical Physics (KITP). V.R. was also partly supported by NASA grant 80NSSC21K1692. C.V. was supported with funding from Type One Energy. Computational resources were provided by the Texas Advanced Computing  Center (TACC) at the University of Texas at Austin and by the NASA High-End Computing (HEC) Program through the NASA Advanced Supercomputing (NAS) Division at Ames Research Center. 
This research also used resources of the National Energy Research Scientific Computing Center, a DOE Office of Science User Facility supported by the Office of Science of the U.S. Department of Energy under Contract No. DE-AC02-05CH11231 using NERSC awards FES-ERCAP0028833 and FES-ERCAP0033257. 
\end{acknowledgments}


\bibliography{references}{}

@ARTICLE{lemoine2024,
       author = {{Lemoine}, Martin and {Murase}, Kohta and {Rieger}, Frank},
        title = "{Nonlinear aspects of stochastic particle acceleration}",
      journal = {\prd},
         year = 2024,
        month = mar,
       volume = {109},
       number = {6},
          eid = {063006},
        pages = {063006},
          doi = {10.1103/PhysRevD.109.063006},
archivePrefix = {arXiv},
       eprint = {2312.04443},
 primaryClass = {astro-ph.HE},
       adsurl = {https://ui.adsabs.harvard.edu/abs/2024PhRvD.109f3006L}
}

@ARTICLE{mbarek2026,
       author = {{Mbarek}, Rostom and {Gro{\v{s}}elj}, Daniel and {Philippov}, Alexander},
        title = "{On the Nonthermal Power Laws in Magnetized Turbulent Plasmas}",
      journal = {\apjl},
         year = 2026,
        month = jun,
       volume = {1003},
       number = {2},
          eid = {L43},
        pages = {L43},
          doi = {10.3847/2041-8213/ae681b},
archivePrefix = {arXiv},
       eprint = {2605.03033},
 primaryClass = {physics.plasm-ph},
       adsurl = {https://ui.adsabs.harvard.edu/abs/2026ApJ..1003L..43M}
}

@ARTICLE{servidio2016,
       author = {{Servidio}, S. and {Haynes}, C.~T. and {Matthaeus}, W.~H. and {Burgess}, D. and {Carbone}, V. and {Veltri}, P.},
        title = "{Explosive Particle Dispersion in Plasma Turbulence}",
      journal = {\prl},
         year = 2016,
        month = aug,
       volume = {117},
       number = {9},
          eid = {095101},
        pages = {095101},
          doi = {10.1103/PhysRevLett.117.095101},
archivePrefix = {arXiv},
       eprint = {1608.01207},
 primaryClass = {physics.plasm-ph},
       adsurl = {https://ui.adsabs.harvard.edu/abs/2016PhRvL.117i5101S}
}

@ARTICLE{grosselj2026,
       author = {{Gro{\v{s}}elj}, Daniel and {Philippov}, Alexander and {Beloborodov}, Andrei M. and {Mushotzky}, Richard},
        title = "{High-energy Emission from Turbulent Electron─Ion Coronae of Accreting Black Holes}",
      journal = {\apj},
         year = 2026,
        month = apr,
       volume = {1001},
       number = {1},
          eid = {64},
        pages = {64},
          doi = {10.3847/1538-4357/ae50fc},
archivePrefix = {arXiv},
       eprint = {2601.00518},
 primaryClass = {astro-ph.HE},
       adsurl = {https://ui.adsabs.harvard.edu/abs/2026ApJ..1001...64G}
}

@ARTICLE{sebastian2025,
       author = {{Sebastian}, Samuel T. and {Comisso}, Luca},
        title = "{Magnetic Field-line Curvature and Its Role in Particle Acceleration by Magnetically Dominated Turbulence}",
      journal = {\apjl},
         year = 2025,
        month = nov,
       volume = {994},
       number = {1},
          eid = {L1},
        pages = {L1},
          doi = {10.3847/2041-8213/ae1696},
archivePrefix = {arXiv},
       eprint = {2510.20628},
 primaryClass = {astro-ph.HE},
       adsurl = {https://ui.adsabs.harvard.edu/abs/2025ApJ...994L...1S}
}

@ARTICLE{wong2025,
       author = {{Wong}, Kai W. and {Zhdankin}, Vladimir and {Uzdensky}, Dmitri A. and {Werner}, Gregory R. and {Begelman}, Mitchell C.},
        title = "{Energy diffusion and advection coefficients in kinetic simulations of relativistic plasma turbulence}",
      journal = {\mnras},
         year = 2025,
        month = oct,
       volume = {543},
       number = {2},
        pages = {1842-1863},
          doi = {10.1093/mnras/staf1589},
archivePrefix = {arXiv},
       eprint = {2502.03042},
 primaryClass = {astro-ph.HE},
       adsurl = {https://ui.adsabs.harvard.edu/abs/2025MNRAS.543.1842W}
}

@ARTICLE{gloeckler1992,
       author = {{Gloeckler}, G. and {Geiss}, J. and {Balsiger}, H. and {Bedini}, P. and {Cain}, J.~C. and {Fischer}, J. and {Fisk}, L.~A. and {Galvin}, A.~B. and {Gliem}, F. and {Hamilton}, D.~C. and {Hollweg}, J.~V. and {Ipavich}, F.~M. and {Joos}, R. and {Livi}, S. and {Lundgren}, R.~A. and {Mall}, U. and {McKenzie}, J.~F. and {Ogilvie}, K.~W. and {Ottens}, F. and {Rieck}, W. and {Tums}, E.~O. and {von Steiger}, R. and {Weiss}, W. and {Wilken}, B.},
        title = "{The Solar Wind Ion Composition Spectrometer}",
      journal = {\aaps},
         year = 1992,
        month = jan,
       volume = {92},
       number = {2},
        pages = {267-289},
       adsurl = {https://ui.adsabs.harvard.edu/abs/1992A&AS...92..267G}
}

@ARTICLE{kolmann2019,
       author = {{Kollmann}, Peter and {Hill}, M.~E. and {McNutt}, Jr., R.~L. and {Brown}, L.~E. and {Allen}, R.~C. and {Clark}, G. and {Andrews}, B. and {Salazar}, N. and {Westlake}, J. and {Romeo}, G. and {Vandegriff}, J. and {Kusterer}, M. and {Smith}, D. and {Jaskulek}, S. and {Decker}, R. and {Cheng}, A.~F. and {Krimigis}, S.~M. and {Lisse}, C.~M. and {Mitchell}, D.~G. and {Weaver}, H.~A. and {Delamere}, P. and {Elliott}, H.~A. and {Fattig}, E. and {Gladstone}, G.~R. and {Valek}, P.~W. and {Weidner}, S. and {Bagenal}, F. and {Hor{\'a}nyi}, M. and {Kammer}, J.~A. and {Kaufmann}, D. and {Olkin}, C.~B. and {Piquette}, M.~R. and {Spencer}, J.~R. and {Steffl}, A.~J. and {Stern}, S.~A. and {Young}, L.~A. and {Ennico}, K. and {Linscott}, I.~R. and {Strobel}, D.~F. and {Summers}, M.~E. and {Szalay}, J.~R.},
        title = "{Suprathermal Ions in the Outer Heliosphere}",
      journal = {\apj},
         year = 2019,
        month = may,
       volume = {876},
       number = {1},
          eid = {46},
        pages = {46},
          doi = {10.3847/1538-4357/ab125f},
       adsurl = {https://ui.adsabs.harvard.edu/abs/2019ApJ...876...46K}
}

@ARTICLE{banik2024,
       author = {{Banik}, Uddipan and {Bhattacharjee}, Amitava and {Sengupta}, Wrick},
        title = "{Universal Nonthermal Power-law Distribution Functions from the Self-consistent Evolution of Collisionless Electrostatic Plasmas}",
      journal = {\apj},
         year = 2024,
        month = dec,
       volume = {977},
       number = {1},
          eid = {91},
        pages = {91},
          doi = {10.3847/1538-4357/ad91a1},
archivePrefix = {arXiv},
       eprint = {2408.07127},
 primaryClass = {astro-ph.SR},
       adsurl = {https://ui.adsabs.harvard.edu/abs/2024ApJ...977...91B}
}

@ARTICLE{mason_g2012,
       author = {{Mason}, G.~M. and {Gloeckler}, G.},
        title = "{Power Law Distributions of Suprathermal Ions in the Quiet Solar Wind}",
      journal = {\ssr},
         year = 2012,
        month = nov,
       volume = {172},
       number = {1-4},
        pages = {241-251},
          doi = {10.1007/s11214-010-9741-0},
       adsurl = {https://ui.adsabs.harvard.edu/abs/2012SSRv..172..241M}
}

@INPROCEEDINGS{gloeckler2008,
       author = {{Gloeckler}, G. and {Fisk}, L.~A. and {Mason}, G.~M. and {Hill}, M.~E.},
        title = "{Formation of Power Law Tail with Spectral Index-5 Inside and Beyond the Heliosphere}",
    booktitle = {Particle Acceleration and Transport in the Heliosphere and Beyond: 7th Annual International AstroPhysics Conference},
         year = 2008,
       editor = {{Li}, Gang and {Hu}, Qiang and {Verkhoglyadova}, Olga and {Zank}, Gary P. and {Lin}, R.~P. and {Luhmann}, J.},
       series = {American Institute of Physics Conference Series},
       volume = {1039},
        month = aug,
    publisher = {AIP},
        pages = {367-374},
          doi = {10.1063/1.2982473},
       adsurl = {https://ui.adsabs.harvard.edu/abs/2008AIPC.1039..367G}
}

@ARTICLE{bykov2001,
       author = {{Bykov}, Andrei M.},
        title = "{Particle Acceleration and Nonthermal Phenomena in Superbubbles}",
      journal = {\ssr},
         year = 2001,
        month = oct,
       volume = {99},
        pages = {317-326},
          doi = {10.1023/A:1013817721725},
       adsurl = {https://ui.adsabs.harvard.edu/abs/2001SSRv...99..317B}
}

@ARTICLE{zhdankin2022,
       author = {{Zhdankin}, Vladimir},
        title = "{Non-thermal particle acceleration from maximum entropy in collisionless plasmas}",
      journal = {Journal of Plasma Physics},
         year = 2022,
        month = jun,
       volume = {88},
       number = {3},
          eid = {175880303},
        pages = {175880303},
          doi = {10.1017/S0022377822000551},
archivePrefix = {arXiv},
       eprint = {2203.13054},
 primaryClass = {astro-ph.HE},
       adsurl = {https://ui.adsabs.harvard.edu/abs/2022JPlPh..88c1703Z}
}

@ARTICLE{ewart2025,
       author = {{Ewart}, Robert J. and {Nastac}, Michael L. and {Bilbao}, Pablo J. and {Silva}, Thales and {Silva}, Lu{\'\i}s O. and {Schekochihin}, Alexander A.},
        title = "{Relaxation to universal non-Maxwellian equilibria in a collisionless plasma}",
      journal = {Proceedings of the National Academy of Science},
         year = 2025,
        month = apr,
       volume = {122},
       number = {17},
          eid = {e2417813122},
        pages = {e2417813122},
          doi = {10.1073/pnas.2417813122},
       adsurl = {https://ui.adsabs.harvard.edu/abs/2025PNAS..12217813E}
}

@ARTICLE{fisk2014,
       author = {{Fisk}, L.~A. and {Gloeckler}, G.},
        title = "{The case for a common spectrum of particles accelerated in the heliosphere: Observations and theory}",
      journal = {Journal of Geophysical Research (Space Physics)},
         year = 2014,
        month = nov,
       volume = {119},
       number = {11},
        pages = {8733-8749},
          doi = {10.1002/2014JA020426},
       adsurl = {https://ui.adsabs.harvard.edu/abs/2014JGRA..119.8733F}
}

@ARTICLE{meringolo2026,
       author = {{Meringolo}, Claudio and {Imbrogno}, Mario and {Cruz-Osorio}, Alejandro and {Servidio}, Sergio and {Rezzolla}, Luciano},
        title = "{Particle-acceleration mechanisms in multispecies relativistic plasmas}",
      journal = {arXiv e-prints},
         year = 2026,
        month = apr,
          eid = {arXiv:2604.06749},
        pages = {arXiv:2604.06749},
          doi = {10.48550/arXiv.2604.06749},
archivePrefix = {arXiv},
       eprint = {2604.06749},
 primaryClass = {astro-ph.HE},
       adsurl = {https://ui.adsabs.harvard.edu/abs/2026arXiv260406749M}
}

@ARTICLE{humphrey2026,
       author = {{Humphrey}, Daniel and {Vega}, Cristian and {Boldyrev}, Stanislav and {Roytershteyn}, Vadim},
        title = "{Particle Acceleration and Pitch-angle Evolution in Relativistic Turbulence}",
      journal = {\apj},
         year = 2026,
        month = feb,
       volume = {997},
       number = {2},
          eid = {276},
        pages = {276},
          doi = {10.3847/1538-4357/ae3173},
archivePrefix = {arXiv},
       eprint = {2603.02342},
 primaryClass = {astro-ph.GA},
       adsurl = {https://ui.adsabs.harvard.edu/abs/2026ApJ...997..276H}
}

@ARTICLE{das2025,
       author = {{Das}, Saikat and {Xu}, Siyao and {N{\"a}ttil{\"a}}, Joonas},
        title = "{Studying mirror acceleration via kinetic simulations of relativistic plasma turbulence}",
      journal = {arXiv e-prints},
         year = 2025,
        month = jun,
          eid = {arXiv:2506.04212},
        pages = {arXiv:2506.04212},
          doi = {10.48550/arXiv.2506.04212},
archivePrefix = {arXiv},
       eprint = {2506.04212},
 primaryClass = {astro-ph.HE},
       adsurl = {https://ui.adsabs.harvard.edu/abs/2025arXiv250604212D}
}

@ARTICLE{abdo2011b,
       author = {{Abdo}, A.~A. and {Ackermann}, M. and {Ajello}, M. and {Baldini}, L. and {Ballet}, J. and {Barbiellini}, G. and {Bastieri}, D. and {Bechtol}, K. and {Bellazzini}, R. and {Berenji}, B. and {Blandford}, R.~D. and {Bloom}, E.~D. and {Bonamente}, E. and {Borgland}, A.~W. and {Bouvier}, A. and {Bregeon}, J. and {Brez}, A. and {Brigida}, M. and {Bruel}, P. and {Buehler}, R. and {Buson}, S. and {Caliandro}, G.~A. and {Cameron}, R.~A. and {Cannon}, A. and {Caraveo}, P.~A. and {Carrigan}, S. and {Casandjian}, J.~M. and {Cavazzuti}, E. and {Cecchi}, C. and {{\c{C}}elik}, {\"O}. and {Charles}, E. and {Chekhtman}, A. and {Chiang}, J. and {Ciprini}, S. and {Claus}, R. and {Cohen-Tanugi}, J. and {Conrad}, J. and {Cutini}, S. and {de Angelis}, A. and {de Palma}, F. and {Dermer}, C.~D. and {Silva}, E. do Couto e. and {Drell}, P.~S. and {Dubois}, R. and {Dumora}, D. and {Escande}, L. and {Favuzzi}, C. and {Fegan}, S.~J. and {Finke}, J. and {Focke}, W.~B. and {Fortin}, P. and {Frailis}, M. and {Fuhrmann}, L. and {Fukazawa}, Y. and {Fukuyama}, T. and {Funk}, S. and {Fusco}, P. and {Gargano}, F. and {Gasparrini}, D. and {Gehrels}, N. and {Georganopoulos}, M. and {Germani}, S. and {Giebels}, B. and {Giglietto}, N. and {Giommi}, P. and {Giordano}, F. and {Giroletti}, M. and {Glanzman}, T. and {Godfrey}, G. and {Grenier}, I.~A. and {Guiriec}, S. and {Hadasch}, D. and {Hayashida}, M. and {Hays}, E. and {Horan}, D. and {Hughes}, R.~E. and {J{\'o}hannesson}, G. and {Johnson}, A.~S. and {Johnson}, W.~N. and {Kadler}, M. and {Kamae}, T. and {Katagiri}, H. and {Kataoka}, J. and {Kn{\"o}dlseder}, J. and {Kuss}, M. and {Lande}, J. and {Latronico}, L. and {Lee}, S. -H. and {Longo}, F. and {Loparco}, F. and {Lott}, B. and {Lovellette}, M.~N. and {Lubrano}, P. and {Madejski}, G.~M. and {Makeev}, A. and {Max-Moerbeck}, W. and {Mazziotta}, M.~N. and {McEnery}, J.~E. and {Mehault}, J. and {Michelson}, P.~F. and {Mitthumsiri}, W. and {Mizuno}, T. and {Monte}, C. and {Monzani}, M.~E. and {Morselli}, A. and {Moskalenko}, I.~V. and {Murgia}, S. and {Nakamori}, T. and {Naumann-Godo}, M. and {Nishino}, S. and {Nolan}, P.~L. and {Norris}, J.~P. and {Nuss}, E. and {Ohsugi}, T. and {Okumura}, A. and {Omodei}, N. and {Orlando}, E. and {Ormes}, J.~F. and {Ozaki}, M. and {Paneque}, D. and {Panetta}, J.~H. and {Parent}, D. and {Pavlidou}, V. and {Pearson}, T.~J. and {Pelassa}, V. and {Pepe}, M. and {Pesce-Rollins}, M. and {Pierbattista}, M. and {Piron}, F. and {Porter}, T.~A. and {Rain{\`o}}, S. and {Rando}, R. and {Razzano}, M. and {Readhead}, A. and {Reimer}, A. and {Reimer}, O. and {Reyes}, L.~C. and {Richards}, J.~L. and {Ritz}, S. and {Roth}, M. and {Sadrozinski}, H.~F. -W. and {Sanchez}, D. and {Sander}, A. and {Sgr{\`o}}, C. and {Siskind}, E.~J. and {Smith}, P.~D. and {Spandre}, G. and {Spinelli}, P. and {Stawarz}, {\L}. and {Stevenson}, M. and {Strickman}, M.~S. and {Suson}, D.~J. and {Takahashi}, H. and {Takahashi}, T. and {Tanaka}, T. and {Thayer}, J.~G. and {Thayer}, J.~B. and {Thompson}, D.~J. and {Tibaldo}, L. and {Torres}, D.~F. and {Tosti}, G. and {Tramacere}, A. and {Troja}, E. and {Usher}, T.~L. and {Vandenbroucke}, J. and {Vasileiou}, V. and {Vianello}, G. and {Vilchez}, N. and {Vitale}, V. and {Waite}, A.~P. and {Wang}, P. and {Wehrle}, A.~E. and {Winer}, B.~L. and {Wood}, K.~S. and {Yang}, Z. and {Yatsu}, Y. and {Ylinen}, T. and {Zensus}, J.~A. and {Ziegler}, M. and {Fermi LAT Collaboration} and {Aleksi{\'c}}, J. and {Antonelli}, L.~A. and {Antoranz}, P. and {Backes}, M. and {Barrio}, J.~A. and {Becerra Gonz{\'a}lez}, J. and {Bednarek}, W. and {Berdyugin}, A. and {Berger}, K. and {Bernardini}, E. and {Biland}, A. and {Blanch}, O. and {Bock}, R.~K. and {Boller}, A. and {Bonnoli}, G. and {Bordas}, P. and {Borla Tridon}, D. and {Bosch-Ramon}, V. and {Bose}, D. and {Braun}, I.},
        title = "{Fermi Large Area Telescope Observations of Markarian 421: The Missing Piece of its Spectral Energy Distribution}",
      journal = {\apj},
         year = 2011,
        month = aug,
       volume = {736},
       number = {2},
          eid = {131},
        pages = {131},
          doi = {10.1088/0004-637X/736/2/131},
archivePrefix = {arXiv},
       eprint = {1106.1348},
 primaryClass = {astro-ph.HE},
       adsurl = {https://ui.adsabs.harvard.edu/abs/2011ApJ...736..131A}
}

@ARTICLE{abdo2011a,
       author = {{Abdo}, A.~A. and {Ackermann}, M. and {Ajello}, M. and {Allafort}, A. and {Baldini}, L. and {Ballet}, J. and {Barbiellini}, G. and {Baring}, M.~G. and {Bastieri}, D. and {Bechtol}, K. and {Bellazzini}, R. and {Berenji}, B. and {Blandford}, R.~D. and {Bloom}, E.~D. and {Bonamente}, E. and {Borgland}, A.~W. and {Bouvier}, A. and {Brandt}, T.~J. and {Bregeon}, J. and {Brez}, A. and {Brigida}, M. and {Bruel}, P. and {Buehler}, R. and {Buson}, S. and {Caliandro}, G.~A. and {Cameron}, R.~A. and {Cannon}, A. and {Caraveo}, P.~A. and {Carrigan}, S. and {Casandjian}, J.~M. and {Cavazzuti}, E. and {Cecchi}, C. and {{\c{C}}elik}, {\"O}. and {Charles}, E. and {Chekhtman}, A. and {Cheung}, C.~C. and {Chiang}, J. and {Ciprini}, S. and {Claus}, R. and {Cohen-Tanugi}, J. and {Conrad}, J. and {Cutini}, S. and {Dermer}, C.~D. and {de Palma}, F. and {Silva}, E. do Couto e. and {Drell}, P.~S. and {Dubois}, R. and {Dumora}, D. and {Favuzzi}, C. and {Fegan}, S.~J. and {Ferrara}, E.~C. and {Focke}, W.~B. and {Fortin}, P. and {Frailis}, M. and {Fuhrmann}, L. and {Fukazawa}, Y. and {Funk}, S. and {Fusco}, P. and {Gargano}, F. and {Gasparrini}, D. and {Gehrels}, N. and {Germani}, S. and {Giglietto}, N. and {Giordano}, F. and {Giroletti}, M. and {Glanzman}, T. and {Godfrey}, G. and {Grenier}, I.~A. and {Guillemot}, L. and {Guiriec}, S. and {Hayashida}, M. and {Hays}, E. and {Horan}, D. and {Hughes}, R.~E. and {J{\'o}hannesson}, G. and {Johnson}, A.~S. and {Johnson}, W.~N. and {Kadler}, M. and {Kamae}, T. and {Katagiri}, H. and {Kataoka}, J. and {Kn{\"o}dlseder}, J. and {Kuss}, M. and {Lande}, J. and {Latronico}, L. and {Lee}, S. -H. and {Lemoine-Goumard}, M. and {Longo}, F. and {Loparco}, F. and {Lott}, B. and {Lovellette}, M.~N. and {Lubrano}, P. and {Madejski}, G.~M. and {Makeev}, A. and {Max-Moerbeck}, W. and {Mazziotta}, M.~N. and {McEnery}, J.~E. and {Mehault}, J. and {Michelson}, P.~F. and {Mitthumsiri}, W. and {Mizuno}, T. and {Moiseev}, A.~A. and {Monte}, C. and {Monzani}, M.~E. and {Morselli}, A. and {Moskalenko}, I.~V. and {Murgia}, S. and {Naumann-Godo}, M. and {Nishino}, S. and {Nolan}, P.~L. and {Norris}, J.~P. and {Nuss}, E. and {Ohsugi}, T. and {Okumura}, A. and {Omodei}, N. and {Orlando}, E. and {Ormes}, J.~F. and {Paneque}, D. and {Panetta}, J.~H. and {Parent}, D. and {Pavlidou}, V. and {Pearson}, T.~J. and {Pelassa}, V. and {Pepe}, M. and {Pesce-Rollins}, M. and {Piron}, F. and {Porter}, T.~A. and {Rain{\`o}}, S. and {Rando}, R. and {Razzano}, M. and {Readhead}, A. and {Reimer}, A. and {Reimer}, O. and {Richards}, J.~L. and {Ripken}, J. and {Ritz}, S. and {Roth}, M. and {Sadrozinski}, H.~F. -W. and {Sanchez}, D. and {Sander}, A. and {Scargle}, J.~D. and {Sgr{\`o}}, C. and {Siskind}, E.~J. and {Smith}, P.~D. and {Spandre}, G. and {Spinelli}, P. and {Stawarz}, {\L}. and {Stevenson}, M. and {Strickman}, M.~S. and {Sokolovsky}, K.~V. and {Suson}, D.~J. and {Takahashi}, H. and {Takahashi}, T. and {Tanaka}, T. and {Thayer}, J.~B. and {Thayer}, J.~G. and {Thompson}, D.~J. and {Tibaldo}, L. and {Torres}, D.~F. and {Tosti}, G. and {Tramacere}, A. and {Uchiyama}, Y. and {Usher}, T.~L. and {Vandenbroucke}, J. and {Vasileiou}, V. and {Vilchez}, N. and {Vitale}, V. and {Waite}, A.~P. and {Wang}, P. and {Wehrle}, A.~E. and {Winer}, B.~L. and {Wood}, K.~S. and {Yang}, Z. and {Ylinen}, T. and {Zensus}, J.~A. and {Ziegler}, M. and {Fermi LAT Collaboration} and {Aleksi{\'c}}, J. and {Antonelli}, L.~A. and {Antoranz}, P. and {Backes}, M. and {Barrio}, J.~A. and {Becerra Gonz{\'a}lez}, J. and {Bednarek}, W. and {Berdyugin}, A. and {Berger}, K. and {Bernardini}, E. and {Biland}, A. and {Blanch}, O. and {Bock}, R.~K. and {Boller}, A. and {Bonnoli}, G. and {Bordas}, P. and {Borla Tridon}, D. and {Bosch-Ramon}, V. and {Bose}, D. and {Braun}, I. and {Bretz}, T. and {Camara}, M. and {Carmona}, E.},
        title = "{Insights into the High-energy {\ensuremath{\gamma}}-ray Emission of Markarian 501 from Extensive Multifrequency Observations in the Fermi Era}",
      journal = {\apj},
         year = 2011,
        month = feb,
       volume = {727},
       number = {2},
          eid = {129},
        pages = {129},
          doi = {10.1088/0004-637X/727/2/129},
archivePrefix = {arXiv},
       eprint = {1011.5260},
 primaryClass = {astro-ph.HE},
       adsurl = {https://ui.adsabs.harvard.edu/abs/2011ApJ...727..129A}
}

@ARTICLE{meyer2010,
       author = {{Meyer}, M. and {Horns}, D. and {Zechlin}, H. -S.},
        title = "{The Crab Nebula as a standard candle in very high-energy astrophysics}",
      journal = {\aap},
         year = 2010,
        month = nov,
       volume = {523},
          eid = {A2},
        pages = {A2},
          doi = {10.1051/0004-6361/201014108},
archivePrefix = {arXiv},
       eprint = {1008.4524},
 primaryClass = {astro-ph.HE},
       adsurl = {https://ui.adsabs.harvard.edu/abs/2010A&A...523A...2M}
}

@ARTICLE{vega2025,
       author = {{Vega}, Cristian and {Boldyrev}, Stanislav and {Roytershteyn}, Vadim},
        title = "{Anisotropic Particle Acceleration in Alfv{\'e}nic Turbulence}",
      journal = {\apj},
         year = 2025,
        month = jun,
       volume = {985},
       number = {2},
          eid = {231},
        pages = {231},
          doi = {10.3847/1538-4357/add147},
archivePrefix = {arXiv},
       eprint = {2504.04306},
 primaryClass = {physics.plasm-ph},
       adsurl = {https://ui.adsabs.harvard.edu/abs/2025ApJ...985..231V}
}

@ARTICLE{vega2024b,
       author = {{Vega}, Cristian and {Boldyrev}, Stanislav and {Roytershteyn}, Vadim},
        title = "{Particle Acceleration in Relativistic Alfv{\'e}nic Turbulence}",
      journal = {\apj},
         year = 2024,
        month = aug,
       volume = {971},
       number = {1},
          eid = {106},
        pages = {106},
          doi = {10.3847/1538-4357/ad5f8f},
archivePrefix = {arXiv},
       eprint = {2405.07891},
 primaryClass = {physics.plasm-ph},
       adsurl = {https://ui.adsabs.harvard.edu/abs/2024ApJ...971..106V}
}

@ARTICLE{xu2023,
       author = {{Xu}, Siyao and {Lazarian}, Alex},
        title = "{Turbulent Reconnection Acceleration}",
      journal = {\apj},
         year = 2023,
        month = jan,
       volume = {942},
       number = {1},
          eid = {21},
        pages = {21},
          doi = {10.3847/1538-4357/aca32c},
archivePrefix = {arXiv},
       eprint = {2211.08444},
 primaryClass = {physics.plasm-ph},
       adsurl = {https://ui.adsabs.harvard.edu/abs/2023ApJ...942...21X}
}

@ARTICLE{french2023,
       author = {{French}, Omar and {Guo}, Fan and {Zhang}, Qile and {Uzdensky}, Dmitri A.},
        title = "{Particle Injection and Nonthermal Particle Acceleration in Relativistic Magnetic Reconnection}",
      journal = {\apj},
         year = 2023,
        month = may,
       volume = {948},
       number = {1},
          eid = {19},
        pages = {19},
          doi = {10.3847/1538-4357/acb7dd},
archivePrefix = {arXiv},
       eprint = {2210.08358},
 primaryClass = {astro-ph.HE},
       adsurl = {https://ui.adsabs.harvard.edu/abs/2023ApJ...948...19F}
}

@ARTICLE{lemoine2021,
       author = {{Lemoine}, Martin},
        title = "{Particle acceleration in strong MHD turbulence}",
      journal = {\prd},
         year = 2021,
        month = sep,
       volume = {104},
       number = {6},
          eid = {063020},
        pages = {063020},
          doi = {10.1103/PhysRevD.104.063020},
archivePrefix = {arXiv},
       eprint = {2104.08199},
 primaryClass = {astro-ph.HE},
       adsurl = {https://ui.adsabs.harvard.edu/abs/2021PhRvD.104f3020L}
}

@ARTICLE{bresci2022,
       author = {{Bresci}, Virginia and {Lemoine}, Martin and {Gremillet}, Laurent and {Comisso}, Luca and {Sironi}, Lorenzo and {Demidem}, Camilia},
        title = "{Nonresonant particle acceleration in strong turbulence: Comparison to kinetic and MHD simulations}",
      journal = {\prd},
         year = 2022,
        month = jul,
       volume = {106},
       number = {2},
          eid = {023028},
        pages = {023028},
          doi = {10.1103/PhysRevD.106.023028},
archivePrefix = {arXiv},
       eprint = {2206.08380},
 primaryClass = {astro-ph.HE},
       adsurl = {https://ui.adsabs.harvard.edu/abs/2022PhRvD.106b3028B}
}

@BOOK{northrop1963,
       author = {{Northrop}, Theodore, G.},
        title = "{The Adiabatic Motion of Charged Particles (Interscience Publishers, Inc., John Wiley \& Sons, New York)}",
         year = 1963,
          doi = {},
       adsurl = {}
}

@ARTICLE{vega2023,
       author = {{Vega}, Cristian and {Boldyrev}, Stanislav and {Roytershteyn}, Vadim},
        title = "{Spatial Intermittency of Particle Distribution in Relativistic Plasma Turbulence}",
      journal = {\apj},
         year = 2023,
        month = jun,
       volume = {949},
       number = {2},
          eid = {98},
        pages = {98},
          doi = {10.3847/1538-4357/accd73},
archivePrefix = {arXiv},
       eprint = {2304.11000},
 primaryClass = {physics.plasm-ph},
       adsurl = {https://ui.adsabs.harvard.edu/abs/2023ApJ...949...98V}
}

@ARTICLE{dong2022,
       author = {{Dong}, Chuanfei and {Wang}, Liang and {Huang}, Yi-Min and {Comisso}, Luca and {Sandstrom}, Timothy A. and {Bhattacharjee}, Amitava},
        title = "{Reconnection-driven energy cascade in magnetohydrodynamic turbulence}",
      journal = {Science Advances},
         year = 2022,
        month = dec,
       volume = {8},
       number = {49},
          eid = {eabn7627},
        pages = {eabn7627},
          doi = {10.1126/sciadv.abn7627},
archivePrefix = {arXiv},
       eprint = {2210.10736},
 primaryClass = {astro-ph.SR},
       adsurl = {https://ui.adsabs.harvard.edu/abs/2022SciA....8N7627D}
}

@ARTICLE{sironi2014,
       author = {{Sironi}, Lorenzo and {Spitkovsky}, Anatoly},
        title = "{Relativistic Reconnection: An Efficient Source of Non-thermal Particles}",
      journal = {\apjl},
         year = 2014,
        month = mar,
       volume = {783},
       number = {1},
          eid = {L21},
        pages = {L21},
          doi = {10.1088/2041-8205/783/1/L21},
archivePrefix = {arXiv},
       eprint = {1401.5471},
 primaryClass = {astro-ph.HE},
       adsurl = {https://ui.adsabs.harvard.edu/abs/2014ApJ...783L..21S}
}

@ARTICLE{nattila2022,
       author = {{N{\"a}ttil{\"a}}, Joonas and {Beloborodov}, Andrei M.},
        title = "{Heating of Magnetically Dominated Plasma by Alfv{\'e}n-Wave Turbulence}",
      journal = {\prl},
         year = 2022,
        month = feb,
       volume = {128},
       number = {7},
          eid = {075101},
        pages = {075101},
          doi = {10.1103/PhysRevLett.128.075101},
archivePrefix = {arXiv},
       eprint = {2111.15578},
 primaryClass = {astro-ph.HE},
       adsurl = {https://ui.adsabs.harvard.edu/abs/2022PhRvL.128g5101N}
}

@ARTICLE{chernoglazov2021,
       author = {{Chernoglazov}, Alexander and {Ripperda}, Bart and {Philippov}, Alexander},
        title = "{Dynamic Alignment and Plasmoid Formation in Relativistic Magnetohydrodynamic Turbulence}",
      journal = {\apjl},
         year = 2021,
        month = dec,
       volume = {923},
       number = {1},
          eid = {L13},
        pages = {L13},
          doi = {10.3847/2041-8213/ac3afa},
archivePrefix = {arXiv},
       eprint = {2111.08188},
 primaryClass = {astro-ph.HE},
       adsurl = {https://ui.adsabs.harvard.edu/abs/2021ApJ...923L..13C}
}

@ARTICLE{zhdankin2020,
       author = {{Wong}, Kai and {Zhdankin}, Vladimir and {Uzdensky}, Dmitri A. and {Werner}, Gregory R. and {Begelman}, Mitchell C.},
        title = "{First-principles Demonstration of Diffusive-advective Particle Acceleration in Kinetic Simulations of Relativistic Plasma Turbulence}",
      journal = {\apjl},
         year = 2020,
        month = apr,
       volume = {893},
       number = {1},
          eid = {L7},
        pages = {L7},
          doi = {10.3847/2041-8213/ab8122},
archivePrefix = {arXiv},
       eprint = {1901.03439},
 primaryClass = {astro-ph.HE},
       adsurl = {https://ui.adsabs.harvard.edu/abs/2020ApJ...893L...7W}
}

@ARTICLE{vega2022a,
       author = {{Vega}, Cristian and {Boldyrev}, Stanislav and {Roytershteyn}, Vadim and {Medvedev}, Mikhail},
        title = "{Turbulence and Particle Acceleration in a Relativistic Plasma}",
      journal = {\apjl},
         year = 2022,
        month = jan,
       volume = {924},
       number = {1},
          eid = {L19},
        pages = {L19},
          doi = {10.3847/2041-8213/ac441e},
archivePrefix = {arXiv},
       eprint = {2111.04907},
 primaryClass = {physics.plasm-ph},
       adsurl = {https://ui.adsabs.harvard.edu/abs/2022ApJ...924L..19V}
}

@ARTICLE{vega2022b,
       author = {{Vega}, Cristian and {Boldyrev}, Stanislav and {Roytershteyn}, Vadim},
        title = "{Spectra of Magnetic Turbulence in a Relativistic Plasma}",
      journal = {\apjl},
         year = 2022,
        month = may,
       volume = {931},
       number = {1},
          eid = {L10},
        pages = {L10},
          doi = {10.3847/2041-8213/ac6cde},
archivePrefix = {arXiv},
       eprint = {2204.04530},
 primaryClass = {physics.plasm-ph},
       adsurl = {https://ui.adsabs.harvard.edu/abs/2022ApJ...931L..10V}
}

@ARTICLE{pezzi2022,
       author = {{Pezzi}, Oreste and {Blasi}, Pasquale and {Matthaeus}, William H.},
        title = "{Relativistic Particle Transport and Acceleration in Structured Plasma Turbulence}",
      journal = {\apj},
         year = 2022,
        month = mar,
       volume = {928},
       number = {1},
          eid = {25},
        pages = {25},
          doi = {10.3847/1538-4357/ac5332},
archivePrefix = {arXiv},
       eprint = {2112.09555},
 primaryClass = {astro-ph.HE},
       adsurl = {https://ui.adsabs.harvard.edu/abs/2022ApJ...928...25P}
}

@ARTICLE{ergun2020a,
       author = {{Ergun}, R.~E. and {Ahmadi}, N. and {Kromyda}, L. and {Schwartz}, S.~J. and {Chasapis}, A. and {Hoilijoki}, S. and {Wilder}, F.~D. and {Cassak}, P.~A. and {Stawarz}, J.~E. and {Goodrich}, K.~A. and {Turner}, D.~L. and {Pucci}, F. and {Pouquet}, A. and {Matthaeus}, W.~H. and {Drake}, J.~F. and {Hesse}, M. and {Shay}, M.~A. and {Torbert}, R.~B. and {Burch}, J.~L.},
        title = "{Particle Acceleration in Strong Turbulence in the Earth's Magnetotail}",
      journal = {\apj},
         year = 2020,
        month = aug,
       volume = {898},
       number = {2},
          eid = {153},
        pages = {153},
          doi = {10.3847/1538-4357/ab9ab5},
       adsurl = {https://ui.adsabs.harvard.edu/abs/2020ApJ...898..153E}
}

@ARTICLE{trotta2020,
       author = {{Trotta}, Domenico and {Franci}, Luca and {Burgess}, David and {Hellinger}, Petr},
        title = "{Fast Acceleration of Transrelativistic Electrons in Astrophysical Turbulence}",
      journal = {\apj},
         year = 2020,
        month = may,
       volume = {894},
       number = {2},
          eid = {136},
        pages = {136},
          doi = {10.3847/1538-4357/ab873c},
archivePrefix = {arXiv},
       eprint = {1910.11935},
 primaryClass = {physics.plasm-ph},
       adsurl = {https://ui.adsabs.harvard.edu/abs/2020ApJ...894..136T}
}

@ARTICLE{comisso2019,
       author = {{Comisso}, Luca and {Sironi}, Lorenzo},
        title = "{The Interplay of Magnetically Dominated Turbulence and Magnetic Reconnection in Producing Nonthermal Particles}",
      journal = {\apj},
         year = 2019,
        month = dec,
       volume = {886},
       number = {2},
          eid = {122},
        pages = {122},
          doi = {10.3847/1538-4357/ab4c33},
archivePrefix = {arXiv},
       eprint = {1909.01420},
 primaryClass = {astro-ph.HE},
       adsurl = {https://ui.adsabs.harvard.edu/abs/2019ApJ...886..122C}
}

@ARTICLE{drake2013,
       author = {{Drake}, J.~F. and {Swisdak}, M. and {Fermo}, R.},
        title = "{The Power-law Spectra of Energetic Particles during Multi-island Magnetic Reconnection}",
      journal = {\apjl},
         year = 2013,
        month = jan,
       volume = {763},
       number = {1},
          eid = {L5},
        pages = {L5},
          doi = {10.1088/2041-8205/763/1/L5},
archivePrefix = {arXiv},
       eprint = {1210.4830},
 primaryClass = {astro-ph.SR},
       adsurl = {https://ui.adsabs.harvard.edu/abs/2013ApJ...763L...5D}
}
\bibliographystyle{aasjournalv7}



\end{document}